\documentclass[11pt,a4paper,english,nofootinbib,,superscriptaddress]{revtex4}
\usepackage{lmodern}
\usepackage{slashed}

\usepackage[T1]{fontenc}
\usepackage[latin9]{inputenc}
\usepackage{babel}
\usepackage{color}
\usepackage{amsmath}
\usepackage{graphicx}
\usepackage{amssymb}
\usepackage{esint}
\usepackage[unicode=true, pdfusetitle,
 bookmarks=true,bookmarksnumbered=false,bookmarksopen=false,
 breaklinks=false,pdfborder={0 0 1},backref=false,colorlinks=false]
 {hyperref}
\usepackage{amsmath}
\usepackage{amssymb}
\def\b{\begin{equation}} \def\e{\end{equation}}
\def\bd{\begin{displaystyle}} \def\ed{\end{displaystyle}}
\def\ba{\begin{array}} \def\ea{\end{array}}

\def\bee{\begin{enumerate}}
\def\eee{\end{enumerate}}

\def\1{\mbox{I\hspace{-.15em}1}}

\def\R{{\rm I\hspace{-.15em}R}}

\def\b{\begin{equation}}
\def\e{\end{equation}}
\def\bee{\begin{enumerate}}
\def\eee{\end{enumerate}}

\makeatletter
\usepackage{latexsym}\usepackage{bm}

\makeatother

\begin{document}
\title{Conjugate spinor field equation for massless spin-$\frac{3}{2}$ field in de Sitter ambient space }

\author{S. Falahi
}
\email{sfalahi27@gmail.com} \affiliation{Department of
Physics,Kermanshah Branch, Islamic Azad University,\\
Kermanshah, Iran}
\author{S. Parsamehr}
\email{sajadparsamehr@iauksh.ac.ir} \affiliation{Department of
Physics,Kermanshah Branch, Islamic Azad University,\\
Kermanshah, Iran}

\date{\today}

\begin{abstract}

\noindent \hspace{0.35cm}
The quantum field theory is expanding in de Sitter space-time. The ambient space formalism let us to develop the quantum field theory in a careful mathematical framework. Using the gauge-covariant derivative in the de Sitter ambient space, the gauge invariant Lagrangian density has been found.
In this paper, the equations of the conjugate spinor for massless spin-$\frac{3}{2}$ gravitational field is obtained by Euler-Lagrange equation. Finally, these equations can be written in terms of the operator of the Casimir group de Sitter.
\end{abstract}

\maketitle
\section{Introduction}
The experimental data and observations show that the universe is expanding with at constant positive acceleration, i.e. the space-time is curved\cite{r1,r2,r3,r4,r5,r6}.
Since the simplest curved space-time that correspondence to these observations is de Sitter space-time and this space-time has maximal symmetry, the group of ten parametric SO (1,4) is the kinematic group of the de Sitter space. Therefore, quantum field theory and gauge theory are investigated in this space-time\cite{qft,qft1,qft2,qft3,qft4,qft5,qft6,qft7,qft8}.\\
 Supersymmetry has been introduced as one of the fundamental principles of all the efforts to achieve  the grand unified theory. It acts in such a way that the relationship between the bosons (integer-valued spin) and the fermions (half-integer spin) is established. In the supersymmetry, all the particles have a partner. For example, for a graviton ( gravity-carrier particle with a spin-2), a partner with a spin-$\frac{3}{2}$ called gravitino can be considered.
With the development of quantum field theory, gauge theories and their very successful results, everyone began to think about quantization and gauge everything. As far as gauge theories are concerned, it should be considered that the structure of a particular symmetry called gauge symmetry, should remain invariant.
Quantum gravity theories that use supersymmetry are called supergravity and seek to unify the gravitational interaction with other fundamental interactions. It should be noted that supergravity is a local supersymmetry theory\cite{s1}. Therefore, gauge theory can be extended to gravity\cite{s2}.\\
To understand physical systems, we must obtain the equation of motion of physical quantities or equally system's Lagrangian.
In Lagrangian mechanics, changes in a physical system are described through solving the Euler-Lagrange equation for that system's behavior. The spin-$\frac{3}{2}$ field equation was first introduced by William Rarita and Julian Schwinger in 1941 using the Euler Lagrange method\cite{rs}.\\
In the previous work\cite{qft6}, we introduced Lagrangian for massless spin-$\frac{3}{2}$ field in de Sitter space-time. Due to the complex and difficult calculations, we intend to present more details of the calculations in this article . The notation of ambient space is briefly reviewed in Section II. In Section III, through this notation, we calculate the conjugate spinor field equation also the equation of this field is invariant. A brief conclusion is presented in Section IV. Finally, details of mathematical calculations are given in two appendices.


\section{notations}

It has been discovered today that the universe is accelerating, with a small, but non-zero and positive cosmological constant, Therefore, it can be concluded that the shape and geometry of the universe is curved. So, in the first-order approximation, we can use  the de Sitter space-time for its. This space-time is a 4-dimensional hyperboloid that can be embedded in a Minkowski 5-dimensional space-time\cite{qft7,qft8}:
\b \label{dSs} X_H=\{x \in \R^5| \; \; x \cdot x=\eta_{\alpha\beta} x^\alpha
 x^\beta =-H^{-2}\},\;\; \alpha,\beta=0,1,2,3,4, \e
where  $\eta_{\alpha\beta}=$diag$(1,-1,-1,-1,-1)$ and H is the Hubble
parameter.
The metric is defined as follows:\b \label
{dsmet}  ds^2=\eta_{\alpha\beta}dx^{\alpha}dx^{\beta}|_{x^2=-H^{-2}}=
g_{\mu\nu}^{dS}dX^{\mu}dX^{\nu},\;\; \mu=0,1,2,3,\e
$X^\mu$ is the four components of the space-time coordinates in a system of intrinsic coordinates on a hyperboloid and $x^\alpha$ is the five dimensional Minkowski space-time(de Sitter ambient space).
Two Casimir operators of the group include:
 \b \label{casi1} Q^{(1)}=-\frac{1}{2}L_{\alpha\beta}L^{\alpha\beta},\;\;  \alpha, \beta=0,1,2,3,4, \e
      \b Q^{(2)}=-W_\alpha W^\alpha\;\;,\;\;W_\alpha =\frac{1}{8}
      \epsilon_{\alpha\beta\gamma\delta\eta} L^{\beta\gamma}L^{\delta\eta},\e
			 $\epsilon_{\alpha\beta\gamma\delta\eta}$ is an anti-symmetric tensor, and $L_{\alpha\beta}$ are ten infinitismal generators in de Sitter space. They can be written as a linear combination:
			$L_{\alpha\beta}=M_{\alpha\beta}+S_{\alpha\beta}$.
			Where $M_{\alpha\beta}$ is the orbital part and $S_{\alpha\beta}$ is the spinoral part. In this formalism, the space $M_{\alpha\beta}$ is represented as follow:
			\b \label{genm} M_{\alpha \beta}=-i(x_\alpha \partial_\beta-x_\beta
      \partial_\alpha)=-i(x_\alpha\partial^\top_\beta-x_\beta
        \partial^\top_\alpha),\e
				where $\partial^\top_\beta=\theta_\beta^{\;\;\alpha}\partial_\alpha$ is the transverse derivative ($x.\partial^\top=0$) and $\theta_{\alpha\beta}=\eta_{\alpha\beta}+H^2x_\alpha x_\beta$ is the projection tensor on de sitter hyperboloid. For half-integer spin fields  $s = l + 1/2$, the spinoral part is defined as:
\begin{equation}
S_{\alpha\beta}^{(s)}=S_{\alpha\beta}^{(l)}+S_{\alpha\beta}^{(\frac{1}{2})},
\end{equation}
where $S_{\alpha\beta}$for spin $\frac{1}{2}$field is:
\b
S_{\alpha\beta}=-\frac{i}{4}[\gamma_{\alpha},\gamma_{\beta}],\e
the $\gamma$-matrices have to satisfy following relation:
\b\{
{\gamma^{\alpha},\gamma^{\beta}}
\}=2\eta^{\alpha\beta}\mathbb{I}.\e
 A proper display for them is \cite{qft7,qft8}:
$$ \gamma^0=\left( \begin{array}{clcr} \mathbb{I} & \;\;0 \\ 0 &-\mathbb{I} \\ \end{array} \right)
      ,\gamma^4=\left( \begin{array}{clcr} 0 & \mathbb{I} \\ -\mathbb{I} &0 \\ \end{array} \right) ,  \label{gammam}
   \gamma^1=\left( \begin{array}{clcr} 0 & i\sigma^1 \\ i\sigma^1 &0 \\
    \end{array} \right)
   ,\gamma^2=\left( \begin{array}{clcr} 0 & -i\sigma^2 \\ -i\sigma^2 &0 \\
      \end{array} \right)
   , \gamma^3=\left( \begin{array}{clcr} 0 & i\sigma^3 \\ i\sigma^3 &0 \\
      \end{array} \right),$$ \b\gamma^{\alpha\dagger}=\gamma^{0}\gamma^{\alpha}\gamma^{0} \qquad  (\gamma^4)^2=-1 \qquad (\gamma^0)^2=1 \e
			where $\mathbb{I}$ is  unit $2\times 2$ matrix and $\sigma^i$ are the pauli matrices.
It should be noted that for massless spin-$\frac{3}{2}$ field $Q^{(1)}_\frac{3}{2}$ is\cite{qft6}:
\begin{equation}
Q_{\frac{3}{2}}^{(1)}\Psi_\alpha=Q_0^{(1)}\Psi_\alpha+\not{x}\not{\partial}^\top\Psi_\alpha
+2x_\alpha\partial^\top\cdot\Psi-\frac{11}{2}\Psi_\alpha+\gamma_\alpha
\not{\Psi},
\end{equation}
$Q_0^{(1)}=-\partial_\alpha^\top\partial^{\alpha\top}$ is the "scalar" Casimir operator.


\section{conjugate spinor field equation for massless spin-$\frac{3}{2}$ field }

It is believed that the gauge theory is the basis of fundamental particle interactions. The Lagrangian for  massless spin-$\frac{3}{2}$ field is peresented, in the linear approximation, by using the gauge theory and defining the gauge covariant derivative\cite{qft6}. The vector-spinor field equation $\Psi_\alpha(x)$ is obtained from the usual Euler-Lagrange equations. This Lagrangian is invariant under the  gauge transformation (see more details [\cite{epjp}]):
\begin{equation}\label{A.1}
{\cal L}=\left(\tilde{\nabla}^\top_\alpha\tilde{
\Psi}_\beta-\tilde{\nabla}^\top_\beta\tilde{
\Psi}_\alpha\right)\left( \nabla^{\top\alpha}
\Psi^\beta-\nabla^{\top\beta} \Psi^\alpha\right),
\end{equation}
$\nabla^\top_\alpha$ is a transverse-covariant derivative which is defined to obtain an invariant Lagrangian according to the following equation:
\b \label{cdsa} \nabla^\top_\beta \Psi_{\alpha_1....\alpha_l}\equiv
(\partial^\top_\beta+\gamma^\top_\beta\not
x)
\Psi_{\alpha_1....\alpha_l}-\sum_{n=1}^{l}x_{\alpha_n}\Psi_{\alpha_1..\alpha_{n-1}\beta\alpha_{n+1}..\alpha_l}.\e
also for the conjugate spinor$\tilde{\Psi}_\alpha$:
\b \label{cdsa} \tilde{\nabla}^\top_\beta \tilde{\Psi}_{\alpha_1....\alpha_l}\equiv
\partial^\top_\beta
\tilde{\Psi}_{\alpha_1....\alpha_l}-\sum_{n=1}^{l}x_{\alpha_n}\tilde{\Psi}_{\alpha_1..\alpha_{n-1}\beta\alpha_{n+1}..\alpha_l}.\e
where $\not x=\gamma_\alpha x^\alpha$and $\gamma^\top_\alpha=\theta_{\alpha}^{\beta}\gamma_\beta$.
The above equations are specifically designed for our calculations:
\begin{equation}
\nabla^\top_\alpha \Psi_\beta= \partial^\top_\alpha \Psi_\beta+ \gamma^\top_\alpha \not x  \Psi_\beta -x_\beta \Psi_\alpha.
\end{equation}
\begin{equation}\tilde{\nabla}^\top_\beta \tilde{\Psi}_\alpha=\partial^\top_\beta
\tilde{\Psi}_\alpha-x_{\alpha}\tilde{\Psi}_\beta
\end{equation}

Now we want to obtain the field equation for the conjugate spinor($\tilde{\Psi}_\alpha={\Psi}^\dag_\alpha\gamma^0$) by using the Euler-Lagrange equation in the linear approximation. The Euler-Lagrange equation is:
\begin{equation}
\frac{\delta {\cal L}}{\delta\tilde{
\Psi}_m}-\partial_{l}^{\top}\frac{\delta {\cal
L}}{\delta(\partial^{\top}_{ l}\tilde{ \Psi}_m)}=0,
\end{equation}
First, we extend each terms of the equation (III.11):
\begin{equation}
A=\left( \nabla^{\top\alpha} \Psi^\beta-\nabla^{\top\beta}
\Psi^\alpha\right)=\left(\partial^{\top\alpha}
\Psi^\beta+\gamma^\alpha\not {x}\Psi^\beta -\partial^{\top\beta}
\Psi^\alpha-\gamma^\beta\not {x}\Psi^\alpha \right),
\end{equation}
\begin{equation}
B=\left(\tilde{\nabla}^\top_\alpha\tilde{
\Psi}_\beta-\tilde{\nabla}^\top_\beta\tilde{ \Psi}_\alpha\right)
=\left(\partial_{\alpha}^{\top}\tilde{ \Psi}_\beta-x_\beta\tilde{
\Psi}_\alpha-\partial_{\beta}^{\top}\tilde{
\Psi}_\alpha+x_\alpha\tilde{ \Psi}_\beta\right),
\end{equation}
By use the Euler-Lagrange equation, we consider the following terms:
\begin{equation}
\frac{\delta {\cal L}}{\delta \Psi_
{m}}=(\gamma^\alpha\not {x} \delta_{\beta}^{m}-\gamma^\beta\not {x} \delta_{\alpha}^{m})\left(\tilde{\nabla}^\top_\alpha\tilde{
\Psi}_\beta-\tilde{\nabla}^\top_\beta\tilde{ \Psi}_\alpha\right),
\end{equation}
and
\begin{equation}
\frac{\delta {\cal L}}{\delta(\partial^{\top}_{ l} \Psi_{m})}=
(\delta_{l}^{\alpha}\delta_{m}^{\beta}-\delta_{l}^{\beta}\delta_{m}^{\alpha})\left(\tilde{\nabla}^\top_\alpha\tilde{
\Psi}_\beta-\tilde{\nabla}^\top_\beta\tilde{ \Psi}_\alpha\right),
\end{equation}
if $\beta=m$, then, one obtains:
\begin{equation}
\frac{\delta {\cal L}}{\delta \Psi_
{m}}=\gamma^\alpha\not {x} \left(\tilde{\nabla}^\top_\alpha\tilde{
\Psi}_\beta-\tilde{\nabla}^\top_\beta\tilde{ \Psi}_\alpha\right),
\end{equation}
\begin{equation}
\frac{\delta {\cal L}}{\delta(\partial^{\top}_{ l} \Psi_{m})}=
\delta_{l}^{\alpha}\left(\tilde{\nabla}^\top_\alpha\tilde{
\Psi}_\beta-\tilde{\nabla}^\top_\beta\tilde{ \Psi}_\alpha\right).
\end{equation}
Now, by placing the above expressions in the Euler-Lagrange equation, the field equation is obtained as follows:
\begin{equation}
\gamma^\alpha\not {x} \left(\tilde{\nabla}^\top_\alpha\tilde{
\Psi}_\beta-\tilde{\nabla}^\top_\beta\tilde{ \Psi}_\alpha\right)-\partial^{\top\alpha} \left(\tilde{\nabla}^\top_\alpha\tilde{
\Psi}_\beta-\tilde{\nabla}^\top_\beta\tilde{ \Psi}_\alpha\right)=0,
\end{equation}
this equation can be written in the summarized form:
\begin{equation}
\left(\partial^{\top\alpha}-\gamma^\alpha\not x \right)
\left(\tilde{\nabla}^\top_\alpha\tilde{
\Psi}_\beta-\tilde{\nabla}^\top_\beta\tilde{ \Psi}_\alpha\right)=0.
\end{equation}

In the appendix A, the field equations for the conjugate spinor is obtained by using the second order Casimir operator:
\begin{equation}
\left(Q_{\frac{3}{2}}^{(1)}+\frac{5}{2}\right)\tilde
\Psi_\alpha+\partial_{\alpha}^\top\left(\not x\tilde{\not \Psi}
+\partial^\top\cdot\tilde\Psi\right)-2\left(\gamma_\alpha\tilde{\not \Psi}
+\not x\not \partial^\top\tilde\Psi_\alpha-\tilde\Psi_\alpha\right)=0.
\end{equation}

In this here we present the gauge invariant field equation for the conjugate spinor.
Given the definition of $Q_{\frac{3}{2}}^{(1)}$, we rewrite Equation (III.25) as follows:
\begin{equation}
-Q_{0}\tilde{ \Psi}_\beta+\tilde{ \Psi}_\beta-2x_\beta(\partial^{\top}\cdot\tilde{
\Psi})-\partial_{\beta}^{\top}(\partial^{\top}\cdot\tilde{
\Psi})+\not x\not \partial^{\top} \tilde{
\Psi}_\beta-x_\beta\not x \not \tilde{\Psi}-\not x\partial_{\beta}^{\top}\not\tilde{
\Psi}=0
\end{equation}
We show that the above equation is invariant under the following gauge transformation:
\begin{equation}
\tilde{\Psi}_\alpha \longrightarrow
\tilde{\Psi}^g_\alpha=\tilde{\Psi}_\alpha+\partial_{\alpha}^\top
\tilde{\psi}.
\end{equation}
Therefore, equation (III.26) comes as follows:
$$
-Q_{0}(\tilde{\Psi}_\beta+\partial_{\beta}^\top
\tilde{\psi})+\tilde{\Psi}_\beta+\partial_{\beta}^\top
\tilde{\psi}-2x_\beta\partial^{\alpha\top}(\tilde{\Psi}_\alpha+\partial_{\alpha}^\top
\tilde{\psi})$$
\begin{equation}
-\partial_{\beta}^{\top}\partial^{\alpha\top}(\tilde{\Psi}_\alpha+\partial_{\alpha}^\top
\tilde{\psi})+\not x\not \partial^{\top} (\tilde{\Psi}_\beta+\partial_{\beta}^\top
\tilde{\psi})-x_\beta\not x (\tilde{\slashed{\Psi}}+\slashed{\partial}^\top\tilde{\psi})
-\not x\partial_{\beta}^{\top}(\tilde{\slashed{\Psi}}+\slashed{\partial}^\top\tilde{\psi})=0
\end{equation}
After we calculate a little:

$$
\Longrightarrow=-Q_{0}\tilde{\Psi}_\beta+\tilde{\Psi}_\beta-2x_\beta(\partial^{\top}\cdot\tilde{
\Psi})-\partial_{\beta}^{\top}(\partial^{\top}\cdot\tilde{
\Psi})+\not x\not \partial^{\top} \tilde{
\Psi}_\beta-x_\beta\not x \not \tilde{\Psi}-\not x\partial_{\beta}^{\top}\not\tilde{
\Psi}$$
\begin{equation}
\underbrace{-Q_{0}\partial_{\beta}^\top\tilde{\psi}+\partial_{\beta}^\top
\tilde{\psi}-2x_\beta\partial^{\alpha\top}\partial_{\alpha}^{\top}\psi-\partial_{\beta}^{\top}\partial^{\alpha\top}\partial_{\alpha}^{\top}\psi
+\slashed{x}\slashed{\partial}^\top\partial_{\beta}^{\top}\tilde{\psi}-x_\beta\not x \not\slashed{\partial}^\top\tilde{\psi}-\slashed{x}
\partial_{\beta}^{\top}\slashed{\partial}^\top\tilde{\psi}}_{=0}=0
\end{equation}

For a gauge invariant, the latest terms must be zero, We have to prove this. To do this we use the following auxiliary relationships:
\begin{equation}
[\partial^\top_\alpha , \not\partial^\top]=\not x\partial^\top_\alpha-x_\alpha\not\partial^\top 
\end{equation}
\begin{equation}
[\partial^\top_\alpha , Q_0]=-6\partial^\top_\alpha-2(Q_0+4)x_\alpha
\end{equation}
Thus, the latest terms of equation (III.29) are obtained as follows:
$$
-Q_{0}\partial_{\beta}^\top\tilde{\psi}+\partial_{\beta}^\top
\tilde{\psi}-2x_\beta\partial^{\alpha\top}\partial_{\alpha}^{\top}\psi-\partial_{\beta}^{\top}\partial^{\alpha\top}\partial_{\alpha}^{\top}\psi
+\slashed{x}\slashed{\partial}^\top\partial_{\beta}^{\top}\tilde{\psi}-x_\beta\not x \not\slashed{\partial}^\top\tilde{\psi}-\slashed{x}
\partial_{\beta}^{\top}\slashed{\partial}^\top\tilde{\psi}
$$
\begin{equation}
=2x_\beta Q_{0}\tilde{\psi}+\partial_{\beta}^\top
\tilde{\psi}-x_\beta\not x \not\slashed{\partial}^\top\tilde{\psi}-6
\partial_{\beta}^{\top}\tilde{\psi}-2Q_{0}x_\beta\tilde{\psi}-8x_\beta \tilde{\psi}+x_\beta\not x \not\slashed{\partial}^\top\tilde{\psi}
+\partial_{\beta}^{\top}\tilde{\psi}
\end{equation}

After a few of simplification:
\begin{equation}
=2x_\beta Q_{0}\tilde{\psi}-4\partial_{\beta}^\top
\tilde{\psi}-2Q_{0}x_\beta\tilde{\psi}-8x_\beta \tilde{\psi}=2(x_\beta Q_{0}-Q_{0} x_\beta )\tilde{\psi}-4\partial_{\beta}^\top
\tilde{\psi}-8x_\beta \tilde{\psi}
\end{equation}
And finally using an auxiliary relationship $[x_\alpha , Q_0]=2\partial^\top_\alpha+4x_\alpha$
 we have:
\begin{equation}
=2(2\partial^\top_\beta+4x_\beta)\tilde{\psi}-4\partial_{\beta}^\top
\tilde{\psi}-8x_\beta \tilde{\psi}=0
\end{equation}

Therefore, we prove that the equation is the invariant field.


\section{Conclusions}

In order to better understand the evolution of the universe, it is necessary to extend the theory of quantum fields, field interactions, or gauge theory, supersymmetry and supergravity in  the de Sitter space-time.
We studied the conjugate spinor field equation by the Euler-Lagrange equation in the de Sitter ambient space formalism. Furthermore, the field equation in terms of the Casimir operator is obtained also the gauge invariant field equation is presented. These studies could be appropriate for supergravity theory in curved space, which could be presented in the future.

\vspace{0.5cm}

\noindent {\bf{Acknowlegements}}:  I would like to express our heartfelt thank and sincere gratitude to Professor
M.V. Takook for his helpful discussions. This work has been supported by the Islamic Azad University, Kermanshah Branch, Kermanshah, Iran.

\appendix

\section{ the field equation in terms of Casimir operator}
Here we want to show how the field equation is written in terms of the Casimir operator:
\begin{equation}\label{B.4}
\left(\partial^{\top\alpha}-\gamma^\alpha\not x \right)\left(\partial_{\alpha}^{\top}\tilde{ \Psi}_\beta-x_\beta\tilde{
\Psi}_\alpha-\partial_{\beta}^{\top}\tilde{
\Psi}_\alpha+x_\alpha\tilde{ \Psi}_\beta\right)=0,
\end{equation}
\begin{equation}
\underbrace{\partial^{\top\alpha}\left(\partial_{\alpha}^{\top}\tilde{ \Psi}_\beta-x_\beta\tilde{
\Psi}_\alpha-\partial_{\beta}^{\top}\tilde{
\Psi}_\alpha+x_\alpha\tilde{ \Psi}_\beta\right)}_{1}
\underbrace{-\gamma^\alpha\not x \left(\partial_{\alpha}^{\top}\tilde{ \Psi}_\beta-x_\beta\tilde{
\Psi}_\alpha-\partial_{\beta}^{\top}\tilde{
\Psi}_\alpha+x_\alpha\tilde{ \Psi}_\beta\right)}_{2}=0
\end{equation}
We first consider the expression 1:
\begin{equation}
\partial^{\top\alpha}\left(\partial_{\alpha}^{\top}\tilde{ \Psi}_\beta-x_\beta\tilde{
\Psi}_\alpha-\partial_{\beta}^{\top}\tilde{
\Psi}_\alpha+x_\alpha\tilde{ \Psi}_\beta\right)=\partial^{\top\alpha}(\partial_{\alpha}^{\top}\tilde{ \Psi}_\beta)-\partial^{\top\alpha}(x_\beta\tilde{
\Psi}_\alpha)-\partial^{\top\alpha}(\partial_{\beta}^{\top}\tilde{
\Psi}_\alpha)+\partial^{\top\alpha}(x_\alpha\tilde{ \Psi}_\beta)
\end{equation}

\begin{equation*}
=-Q_{0}\tilde{ \Psi}_\beta-(\partial^{\top\alpha}x_\beta)\tilde{
\Psi}_\alpha-x_\beta(\partial^{\top\alpha}\tilde{
\Psi}_\alpha)-\partial^{\top\alpha}(\partial_{\beta}^{\top}\tilde{
\Psi}_\alpha)+4\tilde{ \Psi}_\beta
\end{equation*}

\begin{equation*}
=-Q_{0}\tilde{ \Psi}_\beta-(\delta_{\beta}^{\alpha}+x_{\beta}x^{\alpha})\tilde{
\Psi}_\alpha-x_\beta(\partial^{\top}\cdot\tilde{
\Psi})-\partial^{\top\alpha}(\partial_{\beta}^{\top}\tilde{
\Psi}_\alpha)+4\tilde{ \Psi}_\beta
\end{equation*}

\begin{equation*}
=-Q_{0}\tilde{ \Psi}_\beta-\tilde{ \Psi}_\beta-x_\beta(\partial^{\top}\cdot\tilde{
\Psi})-(\partial_{\beta}^{\top}\partial^{\top\alpha}\tilde{
\Psi}_\alpha+x_\beta\partial^{\top\alpha}\tilde{
\Psi}_\alpha-x^\alpha\partial_{\beta}^{\top}\tilde{
\Psi}_\alpha)+4\tilde{ \Psi}_\beta
\end{equation*}

\begin{equation*}
=-Q_{0}\tilde{ \Psi}_\beta+3\tilde{ \Psi}_\beta-2x_\beta(\partial^{\top}\cdot\tilde{
\Psi})-\partial_{\beta}^{\top}(\partial^{\top}\cdot\tilde{
\Psi})+x^\alpha\partial_{\beta}^{\top}\tilde{
\Psi}_\alpha
\end{equation*}

\begin{equation*}
=-Q_{0}\tilde{ \Psi}_\beta+2\tilde{ \Psi}_\beta-2x_\beta(\partial^{\top}\cdot\tilde{
\Psi})-\partial_{\beta}^{\top}(\partial^{\top}\cdot\tilde{
\Psi})
\end{equation*}

We first consider the expression 2:

\begin{equation}
-\gamma^\alpha\not x \left(\partial_{\alpha}^{\top}\tilde{ \Psi}_\beta-x_\beta\tilde{
\Psi}_\alpha-\partial_{\beta}^{\top}\tilde{
\Psi}_\alpha+x_\alpha\tilde{ \Psi}_\beta\right)
\end{equation}

\begin{equation}
=-(2x^\alpha-\not{x} \gamma^\alpha) \left(\partial_{\alpha}^{\top}\tilde{ \Psi}_\beta-x_\beta\tilde{
\Psi}_\alpha-\partial_{\beta}^{\top}\tilde{
\Psi}_\alpha+x_\alpha\tilde{ \Psi}_\beta\right)
\end{equation}

\begin{equation}
=2x^\alpha\partial_{\beta}^{\top}\tilde{
\Psi}_\alpha-2x^\alpha x_\alpha \tilde{
\Psi}_\beta+\not x\not \partial^{\top} \tilde{
\Psi}_\beta-\not x\gamma^\alpha x_\beta\tilde{
\Psi}_\alpha-\not x\gamma^\alpha\partial_{\beta}^{\top}\tilde{
\Psi}_\alpha +\not x x_\alpha\tilde{
\Psi}_\alpha
\end{equation}

\begin{equation}
=\not x\not \partial^{\top} \tilde{
\Psi}_\beta-x_\beta\not x \not \tilde{\Psi}-\not x\partial_{\beta}^{\top}\not\tilde{
\Psi}-\tilde{\Psi}_\beta
\end{equation}
 in the above calculations, we have used the terms $x\cdot\Psi=0 $and $x\cdot\partial^{\top}=0$.
According to 1 and 2 we can write the equation of motion as follow:
\begin{equation}
\Longrightarrow=-Q_{0}\tilde{ \Psi}_\beta+\tilde{ \Psi}_\beta-2x_\beta(\partial^{\top}\cdot\tilde{
\Psi})-\partial_{\beta}^{\top}(\partial^{\top}\cdot\tilde{
\Psi})+\not x\not \partial^{\top} \tilde{
\Psi}_\beta-x_\beta\not x \not \tilde{\Psi}-\not x\partial_{\beta}^{\top}\not\tilde{
\Psi}=0
\end{equation}
Finally, given $Q_{\frac{3}{2}}^{(1)}$ definition, we have:
\begin{equation}
\left(Q_{\frac{3}{2}}^{(1)}+\frac{5}{2}\right)\tilde
\Psi_\alpha+\partial_{\alpha}^\top\left(\not x\tilde{\not \Psi}
+\partial^\top\cdot\tilde\Psi\right)-2\left(\gamma_\alpha\tilde{\not \Psi}
+\not x\not \partial^\top\tilde\Psi_\alpha-\tilde\Psi_\alpha\right)=0.
\end{equation}

\section{ the auxiliary relationships }

Here are some of the auxiliary relationships used in this article:

\begin{equation}
\begin{array}{llcr}\nonumber
[\partial^\top_\alpha , \partial^\top_\beta]=x_\beta\partial^\top_\alpha-x_\alpha\partial^\top_\beta,&
[\partial^\top_\alpha , x_\beta]=\theta_{\alpha\beta},\\

[x_\alpha , \not\partial^\top]=-\gamma_\alpha^\top,&
[\gamma^\top_{\alpha}, \partial^\top_\alpha]=-4\not x,\\

 [Q_{0} , \not x]=-4\not x-2\not\partial^\top,&
[x_\alpha , Q_0]=2\partial^\top_\alpha+4x_\alpha,\\

[\not x , \not\partial^\top]=4-2\not\partial^\top\not x,&
{\gamma^\top_{\alpha}}=\gamma_{\alpha}+x_{\alpha}x\cdot \gamma,\\

[\not x , \partial^\top_\alpha]=-{\gamma^\top_{\alpha}},&
 [\not x , {\gamma^\top_{\alpha}}]=2x_\alpha-2\gamma_\alpha\not x,\\

[\partial^\top_\alpha , \not\partial^\top]=\not x\partial^\top_\alpha-x_\alpha\not\partial^\top ,&
 [{\gamma^\top_{\alpha}} , \partial^\top_\alpha]=-4\not x,\\

 [\partial^\top_\alpha , Q_0]=-6\partial^\top_\alpha-2(Q_0+4)x_\alpha ,&
 [\not\partial^\top , {\gamma^\top_{\alpha}}]=-2{\gamma^\top_{\alpha}}\not\partial^\top+2\partial^\top_\alpha +{\gamma^\top_{\alpha}}\not x+4x_\alpha,\\

  [Q_0 , {\gamma^\top_{\alpha}}]=-8x_\alpha\not x-2\not x\partial^\top_\alpha-2{\gamma^\top_{\alpha}}-2x_\alpha\not\partial^\top .&
 \\
\end{array}
\end{equation}



\end{document}